\documentclass[aps,prl,superscriptaddress,showpacs,floatfix,twocolumn]{revtex4}
\usepackage{amsmath}
\ifx\pdftexversion\undefined
  \usepackage[dvips]{graphicx}
\else
  \usepackage[pdftex]{graphicx}
\fi

\begin{document}

%%%%%%%%%%%%%%%%%%%%%%%%%%%%%%%%%%%%%%%%%%%%%%%%%%%%%%%%%%%%%%%
% Title
%

\title{Absence of Suppression in Particle Production at Large Transverse 
Momentum in $\sqrt{s_{NN}}$~=~200~GeV d+Au Collisions }

\newcommand{\abilene}{Abilene Christian University, Abilene, TX 79699, USA}
\newcommand{\acadsin}{Institute of Physics, Academia Sinica, Taipei 11529, Taiwan}
\newcommand{\barc}{Bhabha Atomic Research Centre, Bombay 400 085, India}
\newcommand{\bnl}{Brookhaven National Laboratory, Upton, NY 11973-5000, USA}
\newcommand{\caucr}{University of California - Riverside, Riverside, CA 92521, USA}
\newcommand{\ciae}{China Institute of Atomic Energy (CIAE), Beijing, People's Republic of China}
\newcommand{\cns}{Center for Nuclear Study, Graduate School of Science, University of Tokyo, 7-3-1 Hongo, Bunkyo, Tokyo 113-0033, Japan}
\newcommand{\colorado}{University of Colorado, Boulder, CO 80309}
\newcommand{\columbia}{Columbia University, New York, NY 10027 and Nevis Laboratories, Irvington, NY 10533, USA}
\newcommand{\dapnia}{Dapnia, CEA Saclay, F-91191, Gif-sur-Yvette, France}
\newcommand{\debrecen}{Debrecen University, H-4010 Debrecen, Egyetem t{\'e}r 1, Hungary}
\newcommand{\elte}{ELTE, E{\"o}tv{\"o}s Lor{\'a}nd University, H - 1117 Budapest, P{\'a}zm{\'a}ny P. s. 1/A, Hungary}
\newcommand{\fsu}{Florida State University, Tallahassee, FL 32306, USA}
\newcommand{\gsu}{Georgia State University, Atlanta, GA 30303, USA}
\newcommand{\hiroshima}{Hiroshima University, Kagamiyama, Higashi-Hiroshima 739-8526, Japan}
\newcommand{\ihepprot}{Institute for High Energy Physics (IHEP), Protvino, Russia}
\newcommand{\illuiuc}{University of Illinois at Urbana-Champaign, Urbana, IL 61801}
\newcommand{\isu}{Iowa State University, Ames, IA 50011, USA}
\newcommand{\jinrdubna}{Joint Institute for Nuclear Research, 141980 Dubna, Moscow Region, Russia}
\newcommand{\kek}{KEK, High Energy Accelerator Research Organization, Tsukuba-shi, Ibaraki-ken 305-0801, Japan}
\newcommand{\kfki}{KFKI Research Institute for Particle and Nuclear Physics (RMKI), H-1525 Budapest 114, POBox 49, Hungary}
\newcommand{\korea}{Korea University, Seoul, 136-701, Korea}
\newcommand{\kurchatov}{Russian Research Center ``Kurchatov Institute", Moscow, Russia}
\newcommand{\kyoto}{Kyoto University, Kyoto 606, Japan}
\newcommand{\labllr}{Laboratoire Leprince-Ringuet, Ecole Polytechnique, CNRS-IN2P3, Route de Saclay, F-91128, Palaiseau, France}
\newcommand{\lawllnl}{Lawrence Livermore National Laboratory, Livermore, CA 94550, USA}
\newcommand{\losalamos}{Los Alamos National Laboratory, Los Alamos, NM 87545, USA}
\newcommand{\lpc}{LPC, Universit{\'e} Blaise Pascal, CNRS-IN2P3, Clermont-Fd, 63177 Aubiere Cedex, France}
\newcommand{\lund}{Department of Physics, Lund University, Box 118, SE-221 00 Lund, Sweden}
\newcommand{\muenster}{Institut fuer Kernphysik, University of Muenster, D-48149 Muenster, Germany}
\newcommand{\myongji}{Myongji University, Yongin, Kyonggido 449-728, Korea}
\newcommand{\nagasaki}{Nagasaki Institute of Applied Science, Nagasaki-shi, Nagasaki 851-0193, Japan}
\newcommand{\newmex}{University of New Mexico, Albuquerque, NM, USA }
\newcommand{\nmsu}{New Mexico State University, Las Cruces, NM 88003, USA}
\newcommand{\ornl}{Oak Ridge National Laboratory, Oak Ridge, TN 37831, USA}
\newcommand{\orsay}{IPN-Orsay, Universite Paris Sud, CNRS-IN2P3, BP1, F-91406, Orsay, France}
\newcommand{\peking}{Peking University, Beijing, People's Republic of China}
\newcommand{\pnpi}{PNPI, Petersburg Nuclear Physics Institute, Gatchina, Russia}
\newcommand{\riken}{RIKEN (The Institute of Physical and Chemical Research), Wako, Saitama 351-0198, JAPAN}
\newcommand{\rkrbrc}{RIKEN BNL Research Center, Brookhaven National Laboratory, Upton, NY 11973-5000, USA}
\newcommand{\saopaulo}{Universidade de S{\~a}o Paulo, Instituto de F\'{\i}sica, Caixa Postal 66318, S{\~a}o Paulo CEP05315-970, Brazil}
\newcommand{\seoulnat}{System Electronics Laboratory, Seoul National University, Seoul, South Korea}
\newcommand{\stonybrkc}{Chemistry Department, Stony Brook University, Stony Brook, SUNY, NY 11794-3400, USA}
\newcommand{\stonycrkp}{Department of Physics and Astronomy, Stony Brook University, SUNY, Stony Brook, NY 11794, USA}
\newcommand{\subatech}{SUBATECH (Ecole des Mines de Nantes, CNRS-IN2P3, Universit{\'e} de Nantes) BP 20722 - 44307, Nantes, France}
\newcommand{\tenn}{University of Tennessee, Knoxville, TN 37996, USA}
\newcommand{\titech}{Department of Physics, Tokyo Institute of Technology, Tokyo, 152-8551, Japan}
\newcommand{\tsukuba}{Institute of Physics, University of Tsukuba, Tsukuba, Ibaraki 305, Japan}
\newcommand{\vandy}{Vanderbilt University, Nashville, TN 37235, USA}
\newcommand{\waseda}{Waseda University, Advanced Research Institute for Science and Engineering, 17 Kikui-cho, Shinjuku-ku, Tokyo 162-0044, Japan}
\newcommand{\weizmann}{Weizmann Institute, Rehovot 76100, Israel}
\newcommand{\yonsei}{Yonsei University, IPAP, Seoul 120-749, Korea}
\newcommand{\deceased}{\dagger}
\affiliation{\abilene}
\affiliation{\acadsin}
\affiliation{\barc}
\affiliation{\bnl}
\affiliation{\caucr}
\affiliation{\ciae}
\affiliation{\cns}
\affiliation{\colorado}
\affiliation{\columbia}
\affiliation{\dapnia}
\affiliation{\debrecen}
\affiliation{\elte}
\affiliation{\fsu}
\affiliation{\gsu}
\affiliation{\hiroshima}
\affiliation{\ihepprot}
\affiliation{\illuiuc}
\affiliation{\isu}
\affiliation{\jinrdubna}
\affiliation{\kek}
\affiliation{\kfki}
\affiliation{\korea}
\affiliation{\kurchatov}
\affiliation{\kyoto}
\affiliation{\labllr}
\affiliation{\lawllnl}
\affiliation{\losalamos}
\affiliation{\lpc}
\affiliation{\lund}
\affiliation{\muenster}
\affiliation{\myongji}
\affiliation{\nagasaki}
\affiliation{\newmex}
\affiliation{\nmsu}
\affiliation{\ornl}
\affiliation{\orsay}
\affiliation{\peking}
\affiliation{\pnpi}
\affiliation{\riken}
\affiliation{\rkrbrc}
\affiliation{\saopaulo}
\affiliation{\seoulnat}
\affiliation{\stonybrkc}
\affiliation{\stonycrkp}
\affiliation{\subatech}
\affiliation{\tenn}
\affiliation{\titech}
\affiliation{\tsukuba}
\affiliation{\vandy}
\affiliation{\waseda}
\affiliation{\weizmann}
\affiliation{\yonsei}
\author{S.S.~Adler}	\affiliation{\bnl}
\author{S.~Afanasiev}	\affiliation{\jinrdubna}
\author{C.~Aidala}	\affiliation{\columbia}
\author{N.N.~Ajitanand}	\affiliation{\stonybrkc}
\author{Y.~Akiba}	\affiliation{\kek} \affiliation{\riken}
\author{A.~Al-Jamel}	\affiliation{\nmsu}
\author{J.~Alexander}	\affiliation{\stonybrkc}
\author{K.~Aoki}	\affiliation{\kyoto}
\author{L.~Aphecetche}	\affiliation{\subatech}
\author{R.~Armendariz}	\affiliation{\nmsu}
% ...
\collaboration{PHENIX Collaboration} \noaffiliation

\date{\today}        

%%%%%%%%%%%%%%%%%%%%%%%%%%%%%%%%%%%%%%%%%%%%%%%%%%%%%%%%%%%%%%%
% Abstract
%

\begin{abstract}
Transverse momentum spectra of charged hadrons with 
$p_T <  8$ GeV/$c$ and neutral pions with $p_T < 10$ GeV/$c$ 
have been measured at midrapidity 
by the PHENIX experiment at RHIC in d+Au collisions 
at $\sqrt{s_{_{NN}}}$ = 200~GeV. 
The measured yields are compared to those in p+p collisions at the 
same $\sqrt{s_{_{NN}}}$ scaled up by the number of 
underlying nucleon-nucleon
collisions in d+Au. The yield ratio does not show the 
suppression observed in central Au+Au collisions at RHIC. Instead,  
there is a small enhancement in the yield of high momentum particles.

\end{abstract}
\maketitle

%%%%%%%%%%%%%%%%%%%%%%%%%%%%%%%%%%%%%%%%%%%%%%%%%%%%%%%%%%%%%
%NOTES: 
%
%1.  For PRL do not use section headings.
%
%2.  Do not worry about indenting the first line of a paragraph.  Just
%    insert a blank line between paragraphs.  Similarly, if you want 
%    an equation to stay within a paragraph, do not put a blank line
%    before or after the equation.
%
%3.  Do not imbed figures or tables; place them all at the end (see below).
%
%4.  Name all references and use "\cite{refname}" in the text to cite them.
%    (The RevTeX macro will replace this with "[1]" in proper PR style.)
%
%5.  The list of references must be ordered in the same sequence as they
%    occur in the text.
%
%6.  Use our standard aknowldegement below as the last paragraph of your
%    text.  (Yes, it does count toward the length!).
%
%%%%%%%%%%%%%%%%%%%%%%%%%%%%%%%%%%%%%%%%%%%%%%%%%%%%%%%%%%%%%
% general introduction
%
% \marginpar{{\small \em Intro}}
%
%\twocolumn

% Introduction
High transverse momentum ($p_T >$ 2 GeV/$c$) hadrons provide
an excellent probe of the high energy density matter created in relativistic 
heavy ion collisions~\cite{jet-quenching,baier}. They 
%are believed to 
arise from fragmentation of quarks and gluons (partons) scattered 
with large momentum transfer, $Q^2$, in the initial parton-parton
interactions \cite{owens}.
In the absence of medium effects, these
hard scattering yields in nucleus-nucleus 
collisions should scale with the average number of inelastic  
nucleon-nucleon collisions $N_{coll}$ (binary scaling).
%~\cite{glauber}. 
One of the most intriguing observations 
from experiments at the  Relativistic Heavy Ion Collider (RHIC) 
is the large suppression of high $p_T$ neutral pion and charged hadron yields
in central Au+Au collisions 
%at $\sqrt{s_{_{NN}}}$ = 130 GeV and 200 GeV
with respect to p+p results scaled by the number of binary nucleon-nucleon
~collisions\cite{Adcox:2001jp,Adcox:2002pe,Adler:2002xw,ppg014}.
%marginpar{WAZ: Cite PHOBOS *IF* they show a Ncoll-scaled result.}
%Yields in peripheral Au+Au collisions are, however, consistent
%with those measured in p+p. 

Theoretical studies of parton propagation in high density matter suggest 
that partons lose a  significant fraction of their energy through gluon 
bremsstrahlung \cite{jet-quenching,baier}, 
%which reduces 
reducing the parton momentum and 
%depletes 
depleting the yield of high $p_{T}$ hadrons 
\cite{quench-effect,vitev,ina,levai,xnwang,arleo}. 
This is a final-state effect in the spatially extended medium 
created in A+A collisions. Initial state effects include 
nuclear modifications to the parton momentum distributions 
(structure functions), and soft scatterings 
%experienced by 
of the
incoming parton prior to its hard scattering. These should be
present in p+A, d+A and A+A.
%collisions. 
%this might be too long a list
%Alternative 
Interpretations of Au+Au collisions based on initial-state parton 
saturation effects~\cite{dima} or final-state hadronic interactions~\cite{gallmeister}
also predict a considerable suppression of the hadron production at high $p_T$.
It is therefore of paramount interest to determine experimentally the modification, if
any, of high $p_T$ hadron yields due to initial state nuclear effects 
%in the absence of
for a system in which
a hot, dense medium is not produced in the final state.
%. We address this here using data obtained from collisions between deuteron 
%and gold beams.
This Letter reports on charged hadron and $\pi^0$ production at
midrapidity 
%in the laboratory frame, 
obtained by the PHENIX~\cite{nim_phenix} 
experiment at RHIC in d+Au collisions at 
$\sqrt{s_{_{NN}}}$ = 200~GeV. The results are
compared to those in p+p and Au+Au~\cite{ppg014,ppg024,ppg023}
at the same nucleon-nucleon center-of-mass energy.  Similar d+Au 
measurements are also reported in~\cite{STARdAu,PHOBOSdAu}.
% I propose to introduce the various predictions of dAu in the discussion
% rather that as more introduction. OK?

%%%%%%%%%%%%%%%%%%%%%%%%%%%%%%%%%%%%%%%%%%%%%%%%%%%%%%%%%%%%%%%%%%%%%%%%
% Experiment and data analysis
%
%During 2002-2003 (Run-3), the
%RHIC collider delivered d+Au collisions 
%at $\sqrt{s_{_{NN}}}$ = 200~GeV, 
%and the PHENIX experiment
%\marginpar{WAZ: Check how accurately it's 100 GeV or 100 GeV/c.}
%sampled an integrated luminosity of 2.3 nb$^{-1}$. 
The data 
%presented in this paper 
%are obtained from a subset of all available data and 
%include 
were collected under two different trigger conditions. 
Minimum bias events with vertex position along 
the beam axis within  $|z| <$~30 cm were triggered by the Beam-Beam 
Counters (BBC)~\cite{nim_phenix} which cover $|\eta|$ = 3.0-3.9. 
The minimum bias trigger accepts (88 $\pm$ 4\%) of all inelastic 
d+Au collisions that satisfy the vertex condition. A total of 1.2
$\times$ 10$^7$ 
and 1.7 $\times$ 10$^7$ events were analyzed 
%to obtain 
for charged hadron and $\pi^0$ spectra, respectively. 
A second ``photon triggered'' sample, requiring showers above 
2.5 GeV for lead-scintillator (PbSc) and 3.5 GeV for lead-glass 
(PbGl) Electromagnetic 
Calorimeters (EMCal)~\cite{nim_phenix} in addition to the BBC requirement, 
is used to extend the $\pi^0$ measurements to higher $p_T$.
This trigger sampled a total of 1.7 $\times$ 10$^9$ events.   
%As this data sample includes the same
%requirement on BBC and ZDC as the minimum bias triggered events, 
%it has the same collision centrality bias.
%\marginpar{WAZ: Above statement is not clear to me; requiring 
%coincidence with MB trigger provides only *upper bound* on centrality bias ot this trigger.}

Neutral pions were measured by the PHENIX EMCal 
via the $\pi^0 \rightarrow \gamma\gamma$ decay. 
The EMCal covers $|\eta| \leq 0.35$ in pseudorapidity and 
$\Delta \phi = \pi$ in azimuth.  It consists of 6 PbSc
and 2 PbGl sectors, each covering $\pi/8$ in azimuth. 
The data from PbSc and PbGl were analyzed separately.
The energy calibration 
%for the EMCal 
is  obtained from beam tests, cosmic rays,
minimum ionizing energy peaks of charged hadrons, and
the energy/momentum ratio of electrons identified with
the Ring Imaging $\check{\rm C}$erenkov Detector (RICH).
The combined uncertainty on the energy scale and linearity 
is $\le$1.5\%, determined from the
response to identified electrons, and confirmed by the positions
and widths of the observed $\pi^0$ mass peaks.

Photon-like energy clusters in the calorimeter are selected via 
shower profile cuts.  The invariant mass for all photon pairs with energy 
asymmetry $|E_1 - E_2|/(E_1+E_2) < 0.7$ is calculated and binned in $p_T$.  
The $\pi^0$ yield in each $p_{T}$ bin is determined by integrating 
the background subtracted two-photon invariant mass distribution 
\cite{ppg014}; the background is determined from mixed events. The
peak-to-background ratio increases from $\sim 0.3$ at $p_T$=1.25 GeV/$c$
to more than 5 above 4.25 GeV/$c$.  The raw  $\pi^0$ spectra are 
corrected by Monte Carlo simulations for trigger efficiency,
acceptance, and for $\pi^0$ reconstruction
efficiency including dead areas, effects of energy resolution,
photon identification cuts and peak extraction window.
Finally, the yields are corrected to the center of 
the $p_T$ bin using the observed slope of the spectrum.
Below $p_T$= 5 GeV/$c$ the yields are determined from the minimum
 bias data
sample while above 5 GeV/$c$ the photon triggered sample is used. 	
%The overall $\pi^0$ reconstruction efficiency at high $p_T$
%is $\sim$76\%. 
The main sources of systematic errors 
are listed in Table~I.
The final systematic errors on the spectra are 
10 to 16\%, increasing with $p_T$.

%%%%%%%%%%%%%%table I
\begin{table}
\caption{\label{tab:syserrneutral}
Systematic errors in percent on $\pi^0$ invariant yields
	for PbSc (PbGl), as a function of $p_T$ (in GeV/$c$). 
%All errors are given in percent and represent $1 \sigma$ ranges. 
%         The error sum for a given $p_T$ column is the quadratic
%         sum of the $p_T$-dependent errors given in that column
%         and the $p_T$-independent errors. The errors are given for 
%	 PbSc(PbGl).  
	 There are 3 categories of uncertainty:
%The errors are categorized by type: 
	 Type {\bf A} is a
	 point-to-point error uncorrelated between $p_T$
	 bins, type {\bf B} is $p_T$ correlated, all points move in
	 the same direction but not by the same factor, while in type {\bf C} 
%is a normalization error in which 
	all points move by the same factor independent of $p_T$.}
\begin{ruledtabular}
\begin{tabular}{ccccc}
    			& type 	& $p_T$=2 & $p_T$=6 & $p_T$=10  \\\hline 
peak extraction		& \bf A & 5.0(5.0) & 5.0(5.0) & 5.0(5.0)  \\
geom. accept.		& \bf B & 3.0(3.0) & 2.0(2.0) & 2.0(2.0)  \\
$\pi^0$ reconstr.eff.  & \bf B & 4.0(4.0) & 4.0(4.0) & 4.5(4.5)  \\
energy scale		& \bf B & 4.0(4.0) & 9.0(9.0) & 11.0(11.0)\\
trigger eff.		& \bf B &  ---     & 5.0(10.0)&  3.0(3.0) \\
trigger norm. 		& \bf C &  ---     & 5.0(5.0) & 5.0(5.0)  \\
conversion corr.        & \bf C & 2.8(2.8) & 2.8(2.8) & 2.8(2.8)  \\\hline
total error 		&       & 8.6(8.6)   & 14(16)   &15(15)
\end{tabular}
\end{ruledtabular}
\end{table}

Charged particles are reconstructed using a drift chamber (DC) followed
by two layers of multiwire proportional chambers with pad readout
(PC1, PC3)~\cite{nim_phenix}. In this analysis 
tracks were reconstructed over a restricted pseudorapidity range
$|\eta| < 0.18$.
%the tracking detectors 
%cover a solid angle of $\mid\eta\mid < 0.18$ and $\Delta \phi = \pi$. 
Pattern recognition in the DC is based on a combinatorial
Hough transform in the track bend plane, while the polar angle is
determined by PC1 and the location of the collision vertex along the
beam direction~\cite{mitchell}. The vertex was
constrained to be within $|z|<$18 cm for analysis of the charged
tracks. The track reconstruction
efficiency is approximately 98\%, independent of $p_T$. Particle momenta are
measured with a resolution $\delta p/p = 0.7\% \oplus 1.1\%p$~(GeV/$c$). 
%\marginpar{WAZ: Scale known how? Particle masses?}
The momentum scale is known to 0.7\%, from the reconstructed
proton mass using the time-of-flight. 
A confirmation hit is required in PC3, located at a
radius of 5 m, within a 2.5$\sigma$ matching window to eliminate
most albedo, conversions, and decays. The remaining background
above $p_T$=5 GeV/$c$ is subtracted statistically using 
%the PC3 matching distributions measured for 
identified conversions
and weak decays (primarily kaons) \cite{ppg023}. 

Corrections to the charged particle spectrum for geometrical acceptance, decays in flight, reconstruction
efficiency, and momentum resolution are determined using a single-particle
GEANT Monte Carlo simulation. All analysis steps, including the outer
detector matching cuts are applied consistently in simulation and data.
As the DC and PC occupancies in d+Au 
%collisions 
are low, no 
multiplicity-dependent occupancy corrections are required. The yield 
was corrected to the center of the $p_T$ bin.
%, as the spectra fall steeply.
Table~II lists the source and magnitude of each contribution
to the systematic uncertainties on the charged
particle spectra.

%%%%%%%%%%%%%%%%table II
\begin{table}
\caption{\label{tab:syserrcharged}
Systematic errors in percent on the $(h^+ + h^-)/2$ invariant yields.
%	 All errors are given in percent and represent $1 \sigma$ ranges.
	 Error types are as in Table~I. }

\begin{ruledtabular}
\begin{tabular}{cccccc}
                  & type  & $p_T<$4 & $p_T$=4 & $p_T$=6 & $p_T$=7 \\ \hline 
PC3 match         & \bf C &  2.4     & 2.4    & 2.4   & 2.4   \\
geometr. acc.     & \bf C &  2.9     & 2.9    & 2.9   & 2.9   \\
Monte Carlo corr. & \bf C & 3.7    & 3.7    & 3.7   & 3.7   \\
mom. resolution   & \bf B & --       & $<$0.5 & 0.6   & 1.2   \\
mom. scale        & \bf B & $<$3.2   & 3.2    & 3.5   & 3.7   \\
backgd. subtrac.  & \bf B & $<$0.5   & 0.6    & 3.8   & 8.2   \\
$\pi$ oversubtrac.& \bf B &  --      & --     & 1.8   & 4.7   \\\hline
Total error       &       & 6.2      & 6.2    & 7.6   & 11.5  \\
\end{tabular}
\end{ruledtabular}
\end{table}

%%%%%%%%%%%%%%%%%%%%%%%%%%%%%%%%%%%%%%%%%%%%%%%%%%%%%%%%%%%%%%%%%%%%%%%%
% Results and discussion
%
%%%%%%%%%%%%%table III
\begin{table}
\caption{\label{tab:syserrref}
Systematic errors on the p+p reference spectrum.
}
\begin{ruledtabular}
\begin{tabular}{cccccc}
                  & type  & $p_T<$2 & $p_T$=4 & $p_T$=6 & $p_T$=10 \\ \hline 
normalization		& \bf C & 10	& 10   & 10   & 10  \\
residual syst.		& \bf B & 3	& 5   & 7   & 9 \\
charged ($\pi^0$) only\footnotemark[1] 
		        & \bf B(C) &  8(5) & 7(5) & 9(5) & 20(5) \\
\end{tabular}
\end{ruledtabular}
\footnotetext[1]{Errors affect only the charged (neutral)
analysis, and include h/$\pi_0$, and fits to reference (point by
point pp) errors.}
\end{table}

\begin{figure}[tb]
\center
\includegraphics[width=1.0\linewidth]{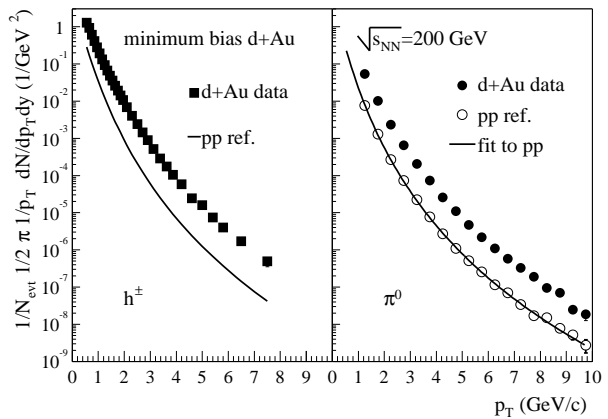}
%\centerline{\epsfig{file=fig1.eps,width=1.0\linewidth}}
\caption[]{\label{fig:pt_spectra} 
Midrapidity $p_T$ spectra for charged hadrons and $\pi^0$.
Total uncertainties are shown. The $\pi^0$ below 5 GeV/$c$ 
are from minimum bias triggered 
events, and above from photon triggered. 
%In addition, the right panel shows 
%the $\pi^0$ spectrum in p+p measured by PHENIX and a 
%power-law fit to the data~\cite{ppg024}. 
%The line on the left panel
%shows the charged hadron reference spectrum in p+p collisions, which
%was derived from the PHENIX $\pi^0$ data as described in the text.
Lines show fits to reference spectra from p+p collisions, and open points
show the p+p $\pi^0$ spectrum measured by PHENIX~\cite{ppg024}. 
}
\end{figure}

The fully corrected $p_T$ distributions  of $\pi^0$'s and 
%inclusive charged particles 
$(h^+ + h^-)/2$ are shown 
%for the minimum bias and photon trcharged hadron iggered data 
in Fig.~\ref{fig:pt_spectra}. 
Each panel shows the reference
spectrum from p+p collisions along with the particle spectrum
from d+Au. The reference 
%spectrum 
for $\pi^0$ is the $\pi^0$ spectrum
measured in p+p 
%collisions 
by PHENIX~\cite{ppg024}, shown on the right panel.
For charged hadrons, the reference is obtained by scaling
%measured $\pi^0$ spectrum in p+p
%\cite{ppg024} 
%is scaled 
up the $\pi^0$ spectrum by the $p_T$-dependent h/$\pi$ ratio
observed at the ISR~\cite{alper} at $\sqrt{s}=23$ to 63 GeV
and measured by PHENIX. 
The value of h/$\pi$ is constant at 1.6 $\pm$ 0.16 
above 1.5 GeV/$c$,
and decreases at lower $p_T$ to a value of $1.25^{+0.12}_{-0.25}$.
Table~III summarizes uncertainties on the reference spectra.
%The systematic uncertaities on the reference are given in Table~III. 
%Bands indicate uncertainties in the reference spectrum,
%arising primarily from the normalization of the PHENIX $\pi^0$ spectrum
%in p+p and uncertainty in the h/$\pi$ ratio.
%Systematic errors are listed in Tables 1 and 2.
%The spectra can be fit with a modified power-law shape 
%$1/(2\pi p_T)d^2\sigma /d\eta dp_T = A/(1 + p_T/p_0)^n$.
%We find for the neutral pion spectrum A = 21.8, $p_0$ = 2.15, and
%n = 11.04, while for the charged spectrum A = 17.8, $p_0$ = 2.46, and n =
%12.7 .
%\marginpar{WAZ: Fit parameters needed here}

%%%%%%%%%%%%%%%%%%%%%%%%%%%%%%%%%%%%%%%% Figure 1.
%%%%%%%%%%%%%%%%%%%%%%%%%%%%%%%%%%%%%%

A standard way to quantify nuclear medium effects on 
high $p_T$ production 
%in nucleus-nucleus collisions 
is 
%provided 
by the 
{\it nuclear modification factor}, which we define for d+Au collisions 
as the ratio of 
%the measured d+A invariant yields to the 
%binary scaled p+p invariant yields:
invariant yield in d+Au to that of p+p, scaled by the number of binary collisions.
\begin{displaymath} 
\nonumber
R_{\rm{dA}}(p_T)\,=\,\frac{(1/N^{evt}_{\rm{dA}})\,d^2N_{\rm{dA}}/d\eta dp_T}
{\langle N_{coll}\rangle /\sigma_{\rm{pp}}^{inel}\,d^2\sigma_{\rm{pp}}/d\eta dp_T},
%\label{eq:R_dA}
\end{displaymath}
where 
$\langle N_{coll}\rangle$ %in Eq. (\ref{eq:R_dA})
is the average number of inelastic nucleon-nucleon (NN) collisions per event in the 
minimum bias collisions, and
$\langle N_{coll} \rangle /\sigma_{\rm{pp}}^{inel}$ is the nuclear
overlap function $\langle T_A(b) \rangle$.
Using a Glauber model~\cite{mult}
and simulation of the BBC, $\langle N_{coll} \rangle$ is
8.5 $\pm$ 0.4 in minimum bias d+Au.
%for the minimum bias trigger sample.
%(Table~\ref{tab:Ncoll}). $R_{dA}(p_T)$ measures the deviation 
%of AA data from the incoherent superposition of NN collisions
%as expected for hard processes with small cross-sections.

\begin{figure}[tb]
\includegraphics[width=1.0\linewidth]{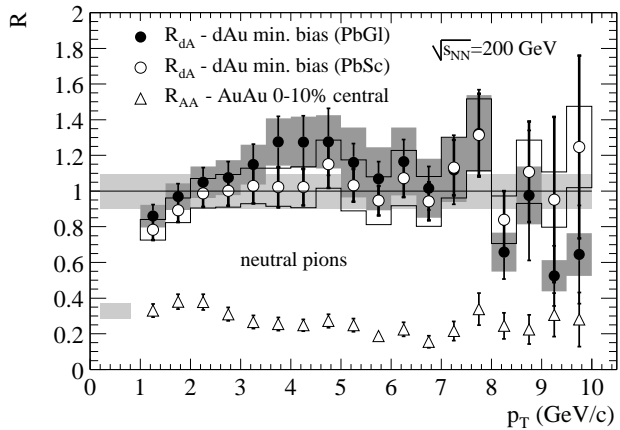}
%\centerline{\epsfig{file=fig2.eps,width=1.0\linewidth}}
\caption{\label{fig:R_AA} 
Nuclear modification factor $R_{\rm{dA}}$ for $\pi^0$ 
%measured separately 
in the PbGl and PbSc calorimeters in minimum bias d+Au.
The bands around the data points show systematic errors which
can vary with $p_T$, while the shaded band around unity indicates
the normalization uncertainty.
%around unity indicates systematic errors common to both data sets,
%while the bands around the data points give the independent systematic 
%uncertainties. 
The nuclear modification factor 
$R_{\rm{AA}}$ in 10\% most central Au+Au collisions is also shown.
}
\end{figure}

\begin{figure}[tb]
%\vspace{-0.9cm}
\includegraphics[width=1.0\linewidth]{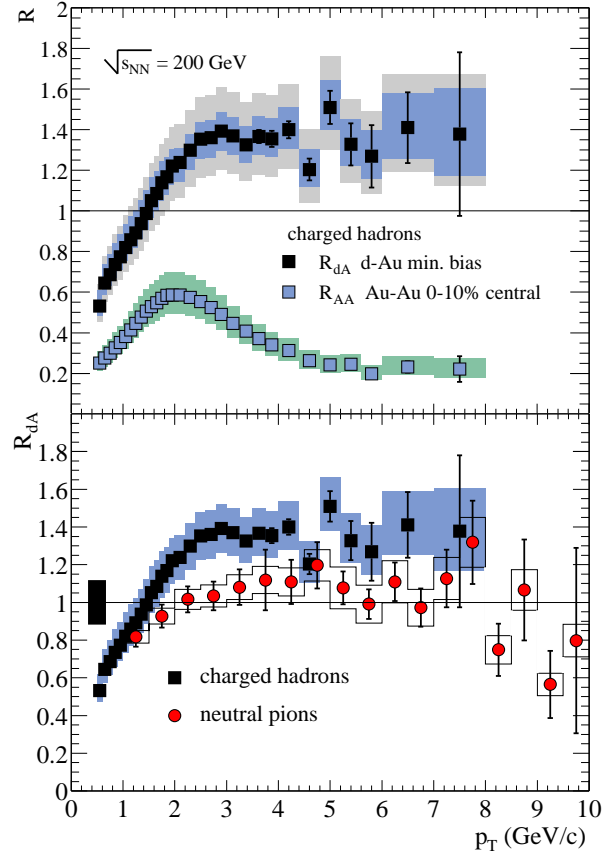}
%\centerline{\epsfig{file=fig3.eps,width=1.0\linewidth}}
%\vspace{0.cm}
\caption{\label{fig:R_AA_vs_dA}
Top: Nuclear modification factor $R_{\rm{dA}}$ for  $(h^+ + h^-)/2$ 
in minimum bias d+Au compared 
to $R_{\rm{AA}}$ in the 10\% most central Au+Au collisions. 
Inner bands show systematic errors which can vary with $p_T$,
and outer bands include also the normalization uncertainty.
%Ratio of $(h^+ + h^-)/2$ yield scaled by the number
%of binary NN collisions, compared to that in the 0--10\% most
%central Au+Au collisions at $\sqrt{s_{_{NN}}}$ = 200~GeV.
Bottom: Comparison of $R_{\rm{dA}}$ for $(h^+ + h^-)/2$ and 
the average of the $\pi^0$ measurements in d+Au. 
%The two d+Au data sets in Fig.~\ref{fig:R_AA} were averaged. 
The bar at the left 
indicates the systematic uncertainty in common for the charged 
and $\pi^0$ measurements.
}
\end{figure}

The ratio $R_{\rm{dA}}$ is plotted separately for $\pi^0$ measured
in the PbGl and in the PbSc calorimeters in Fig.~\ref{fig:R_AA}. The
two analyses are consistent within errors. The data are compared to the
corresponding nuclear modification factor $R_{\rm{AA}}$ obtained from 
central Au+Au collisions.
The top panel of Fig.~\ref{fig:R_AA_vs_dA} shows
$R_{\rm{dA}}$ for inclusive charged particles
$(h^+ + h^-)/2$, again compared with $R_{\rm{AA}}$
observed in central Au+Au collisions, while the
lower panel compares $(h^+ + h^-)/2$ with $\pi^0$.
%, and compared to the results from central Au+Au 
%collisions at $\sqrt{s}$ = 200 GeV. 
In both Figs.~\ref{fig:R_AA} and \ref{fig:R_AA_vs_dA}, 
the uncertainties are plotted as follows: 
error bars represent the quadrature sum of statistical errors and those 
systematic errors which vary point-to-point in $p_T$; 
systematic errors on the absolute yield and the systematic errors
which are correlated point-to-point are shown as bands. 
%The systematic errors
%correspond to one standard deviation uncertainties, under the assumption
%of gaussian distributed errors for the neutral pions and uniform
%distribution of uncertainty for inclusive charged particles.
%\marginpar{WAZ: !! EEK !! Why are pi0's and charged treated differently? This will raise alarm bells without much
%more explanation. What I think is being said is that two different groups used two different
%procedures to derive errors, which were then converted to corresponding 1-sigma values. 
%That's OK, but it reads very funny. Since we don't
%include enough details on how the brackets, ranges, etc. are obtained, 
%I think it's better to leave this out, 
%or to replace it with something like }
All procedures for estimating the various systematic
errors have been adjusted to provide estimates corresponding to 
1$\sigma$ error values.

%%%%%%%%%%%%%%%%%%%%%%%%%%%%%%%%%%%%%%%% Figure 2.
%\vspace{-0.5cm}

%%%%%%%%%%%%%%%%%%%%%%%%%%%%%%%%%%%%%%%% Figure 3.
%\vspace{-0.5cm}

The data clearly indicate that there is no 
%significant 
suppression
of high $p_T$ particles in d+Au collisions. We do, however, observe 
an enhancement in inclusive charged particle production at $p_T >$ 2 GeV/$c$. 
A similar enhancement 
%has been 
was observed in p+A fixed-target experiments
\cite{cronin} and is generally referred to as the ``Cronin effect".
To facilitate comparison of the Cronin effect in 
inclusive charged 
particles and $\pi^0$, the lower part of
Fig.~\ref{fig:R_AA_vs_dA} shows 
all systematic uncertainties common to both analyses in the bar
on the left. It should be noted that this uncertainty must be added 
in quadrature with the bands shown for each curve 
to obtain the 1$\sigma$ allowed range
%maximum and minimum values 
of $R_{\rm{dA}}$ from the data.
The $\pi^0$ data suggest
%are quite similar to the charged $R_{\rm{dA}}$ values, but our data suggest 
a smaller enhancement for 
pions than for inclusive charged particles at $p_T$ = 2-4 GeV/$c$. 
We note that the charged spectrum includes baryons and antibaryons,
which may have a different nuclear enhancement than the 
mesons~\cite{cronin}.

The various models of the suppression observed in Au+Au %collisions
predict a different dependence on $N_{coll}$
in d+Au~\cite{dima,vitev2,boris}. 
%Furthermore, as the Cronin effect
%is often ascribed to multiple scattering in the initial state, the
%dependence of the observed enhancement upon the d+Au 
%collision centrality 
%impact parameter will further test models of the Cronin effect.
Therefore, a second data sample was selected
by requiring observation of a neutron in the Zero-Degree Calorimeter
on the deuteron-going side of PHENIX. This, together with the
requirement of particles entering both Beam-Beam Counters, selects
a class of events in which only the  
proton from the deuteron interacts with the Au nucleus.
The mean number of binary collisions for this sample is calculated with 
the Glauber model to be 3.6 $\pm$ 0.3. 
%The yield of hadrons and $\pi^0$
Particle yields
in this sample have a $<$~5\% uncertainty beyond that of the
minimum bias sample, arising from trigger bias.

\begin{figure}[tb]
\includegraphics[width=1.0\linewidth]{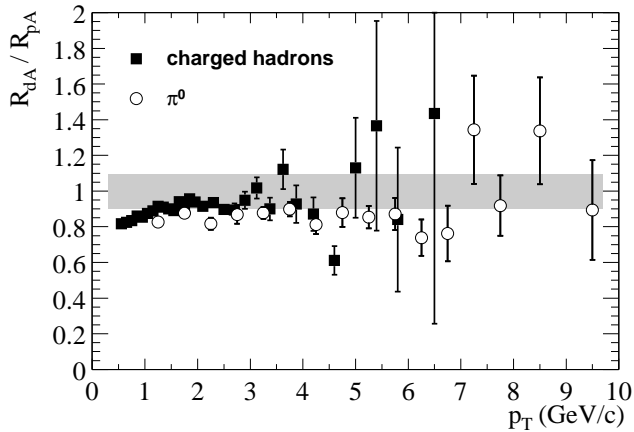}
%\centerline{\epsfig{file=fig4.eps,width=1.0\linewidth}}
\caption{\label{fig:cent_dependence} 
Ratio of $R_{\rm{dA}}$ in minimum bias d+Au and $R_{\rm{pA}}$ from neutron
tagged d+Au collisions for $(h^+ + h^-)/2$ and $\pi^0$.
%The ratio is calculated from the yields normalized 
%to $\langle N_{coll}\rangle$ for each sample;  8.5$\pm$0.4 for minimum 
%bias and  3.6$\pm$0.25 for neutron tagged d+Au collisions. 
%Ratio of $(h^+ + h^-)/2$ yield in minimum bias and
%forward neutron tagged collisions. $<N_{coll}>$ = 8.5 $\pm$ 0.4
%for the former and 3.6 $\pm$ 0.25 for the latter. 
The band shows the uncertainty on the number of binary collisions;
all other systematic errors cancel.
%All systematic errors cancel in the ratio, except the number 
%of binary collisions uncertainty, which is indicated as band.
}
\end{figure}

Fig.~\ref{fig:cent_dependence} shows the ratios of $R_{\rm{dA}}$
in minimum bias d+Au to $R_{\rm{pA}}$ in the neutron tagged sample,
for both $(h^+ + h^-)/2$ and $\pi^0$. 
%spectra normalized to $\langle N_{coll}\rangle$ 
%from minimum bias d+Au to those in
%the neutron tagged p+Au sample, both for 
%$(h^+ + h^-)/2$ and $\pi^0$ spectra. 
%Only statistical errors are shown, as all 
Systematic uncertainties on the spectra
%analyses 
cancel in the ratio; the band around unity shows
the uncertainty on the ratio of
the number of binary collisions in the two samples.
% ; the width of the band indicates the
% uncertainty in the ratio of the number of collisions. 
Average values of $N_{coll}$ are 3.6 per participating proton
in the neutron tagged sample and 8.5 for
1.7 participating nucleons from the deuteron in minimum bias d+Au. 
Given the systematic uncertainties on $N_{coll}$, we cannot
exclude a small centrality dependence
for $p_T > 1$ GeV/$c$. 
%the two samples are consistent with scaling by the number of binary collisions,
It should be noted that the figure also indicates that d+Au collisions
provide a good measure of the physics of p+Au.

%%%%%%%%%%%%%%%%%%%%%%%%%%%%%%%%%%%%%%%% Figure 4.
%\vspace{-0.5cm}

%The presence of the Cronin enhancements in d+Au collisions indicates 
%that the suppression
%observed in the Au+Au data is not an initial state effect arising
%from modifications of structure functions in nuclei. 
%from the binding of the nucleons in nuclei. 
%The data suggest, instead, that suppression 
%of high $p_T$ hadrons is more likely a final state effect of
%the produced medium. The apparent independence 
%of the enhancement on the number of binary nucleon-nucleon collisions 
%in d+Au may pose a challenge to initial-state parton saturation 
%models\cite{dima} and to initial state multiple scattering explanations
%of the Cronin effect.  However, we note that neither of our data samples is
%dominated by very central collisions, where initial state effects
%should be maximal.

The observation of an enhancement of high-$p_T$ hadron production in both 
the minimum bias d+Au and the neutron tagged sample of p+Au 
collisions indicates that the suppression in central Au+Au collisions 
is not an initial state
effect. Nor does it arise from modification of parton structure
functions in nuclei. The data suggest, instead, that the suppression
of high $p_T$ hadrons in Au+Au is more likely a final state effect of the
produced dense medium. 

%%%%%%%%%%%%%%%%%%%%%%%%%%%%%%%%%%%%%%%%%%%%%%%%%%%%%%%%%%%%%%%%%
% Conclusion
%
%\marginpar{{\small \em Concl}}

%In summary, we have presented spectra for charged hadrons and neutral
%pions measured at midrapidity in d+Au collisions in the PHENIX experiment
%at RHIC. Above $p_T \approx$ 2 GeV/$c$, we do not observe the strong
%suppression present in central Au+Au collisions. Instead, we observe
%a slight enhancement, as expected from the Cronin effect in fixed target
%experiments at lower energy. The enhancement is similar for minimum
%bias d+Au collisions and a sample of neutron
%tagged p+Au collisions.
%These results indicate that
%the suppression in Au+Au is more likely an effect of the 
%final, rather than the initial, state of the collision.

%%%%%%%%%%%%%%%%%%%%%%%%%%%%%%%%%%%%%%%%%%%%%%%%%%%%%%%%%%%%
%\section{Acknowledgements}   % Run-3 short form for PRL

We thank the staff of the Collider-Accelerator and Physics
Departments at BNL for their vital contributions.  We acknowledge
support from the Department of Energy and NSF (U.S.A.), MEXT and
JSPS (Japan), CNPq and FAPESP (Brazil), NSFC (China), IN2P3/CNRS
and CEA (France), BMBF, DAAD, and AvH (Germany), OTKA
(Hungary), DAE and DST (India), ISF (Israel), KRF and CHEP (Korea),
RMIST, RAS, and RMAE (Russia), VR and KAW (Sweden), U.S. CRDF for
the FSU, US-Hungarian NSF-OTKA-MTA, and US-Israel BSF.

%REFERENCES:  Use \begin{references} and \end{references}.  Do not use
%             \begin{thebibliography} and \end{thebibliography}.
%             You may either 
%		(a) enter all citations explicitly or 
%               (b) use some "\def" shorthand notations.
%             Our first paper used approach (a) and our second used (b).
%             Here are the two reference lists as examples of how to proceed:

%%%%%% Tables

%FIGURES:  Place all the figures here (after the references) in sequence.

\end{document}